\documentclass[12pt]{article}
\usepackage[letterpaper,margin=1in]{geometry}
\usepackage{graphicx, amsthm, enumitem, tikz, multicol, bbm, bigints, amsfonts, amsmath, mathtools, hyperref} 
\usepackage [autostyle, english = american]{csquotes}
\usepackage[figurename=Fig., labelfont=bf, font=small]{caption}
\usepackage[export]{adjustbox}
\usepackage[skip=1ex, belowskip=2ex]{subcaption}
\usepackage[toc,page]{appendix}
\usepackage{authblk}
\usepackage{colortbl}
\usepackage{xcolor}
\usepackage{pgf}
\usepackage{array}
\usepackage{collcell}
\usepackage{booktabs}
\usepackage[numbers]{natbib}
\usepackage{etoolbox}
\AtBeginEnvironment{thebibliography}{\small}

\usepackage[letterpaper,margin=1in]{geometry}

\newcommand*{\MinNumber}{-10}%
\newcommand*{\MidNumber}{0}%
\newcommand*{\MaxNumber}{10}%

\newcommand{\ApplyGradient}[1]{%
    \ifdim #1 pt < \MidNumber pt
        \pgfmathsetmacro{\PercentColor}{100 - 99.5*(#1-\MinNumber)/(\MidNumber-\MinNumber)}%
        \edef\x{\noexpand\cellcolor{red!\PercentColor}}\x\textcolor{black}{#1}%
    \else
        \pgfmathsetmacro{\PercentColor}{160*(#1-\MidNumber)/(\MaxNumber-\MidNumber)}%
        \edef\x{\noexpand\cellcolor{green!\PercentColor}}\x\textcolor{black}{#1}%
    \fi
}

\newcolumntype{R}{>{\collectcell\ApplyGradient}r<{\endcollectcell}}

\title{Understanding U.S. Racial Segregation Through Persistent Homology}
\author[1]{Ori Friesen}
\author[1]{Lori Ziegelmeier}
\affil[1]{Department of Mathematics, Statistics, and Computer Science, Macalester College, Saint Paul, MN, USA}
\date{May 2024}

\begin{document}

\maketitle
\begin{abstract}
\noindent
    Racial segregation is a widespread social and physical phenomenon present in every city across the United States. Although prevalent nationwide, each city has a unique history of racial segregation, resulting in distinct ``shapes'' of segregation. We use persistent homology, a technique from applied algebraic topology, to investigate whether common patterns of racial segregation exist among U.S. cities. We explore two methods of constructing simplicial complexes that preserve geospatial data, applying them to White, Black, Asian, and Hispanic demographic data from the U.S. census for 112 U.S. cities. Using these methods, we cluster the cities based on their persistence to identify groups with similar segregation ``shapes''. Finally, we apply cluster analysis techniques to explore the characteristics of our clusters. This includes calculating the mean cluster statistics to gain insights into the demographics of each cluster and using the Adjusted Rand Index to compare our results with other clustering methods.
\end{abstract}

\section{Introduction}

The term \textit{racial segregation} refers to the phenomenon where people are physically and socially separated based on their racial and ethnic identity~\cite{delaney2010race}. In the United States, the term carries a much heavier connotation as it relates to the country's history of racial segregation, especially between White and Black communities~\cite{historyofsegregationus}. Racial segregation can take many forms (occupational, educational, social, political, residential, and economic) and is a result of both private practices (called \textit{de facto} segregation) and public policy and law (\textit{de jure} segregation)~\cite{americanapartheid}. While \textit{de jure} racial segregation was outlawed by the Civil Rights Act of 1964, the Voting Rights Act of 1965, and the Fair Housing Act of 1968, we still see many forms of segregation in the present-day, especially residential segregation \cite{coloroflaw}. The long-standing history of residential segregation in the United States means that many communities still experience racial segregation presently. While residential segregation is not the only form that still exists, it is the easiest to discern because the U.S. government collects data on it every ten years through the decennial Census \cite{census_data}. While not explicitly data on residential segregation, the decennial Census attempts to denote the racial and ethnic makeup of every household in the United States, which we can then use to observe larger patterns of dissimilarity between races and ethnicities.

\begin{figure}[htbp]
    \centering
    \setlength{\tabcolsep}{12pt}
    \begin{tabular}{cc}
        \includegraphics[width=0.35\linewidth]{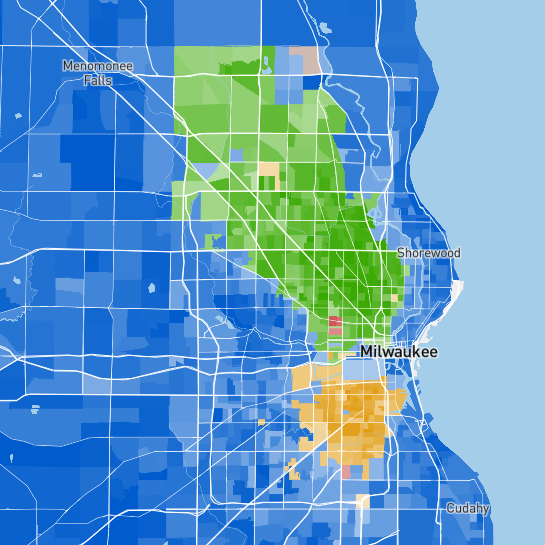}&
        \includegraphics[width=0.35\linewidth]{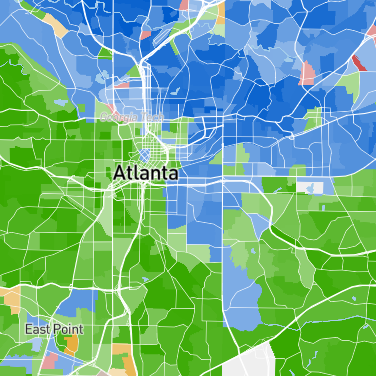}
    \end{tabular}
    \caption{Racial distribution maps of Milwaukee, WI (left), and Atlanta, GA (right), majority White areas depicted in blue, majority Black areas in green, majority Hispanic areas in yellow, and majority Asian areas in red. Source: \cite{bestneighborhood}.}
    \label{fig:intro_cities}
\end{figure}

Because residential segregation, hereafter referred to as segregation, is a phenomenon existing in virtually every U.S. city, questions surrounding how structures of segregation compare across the U.S. are certain to emerge. For example, take the racial distribution maps of Milwaukee and Atlanta in Figure \ref{fig:intro_cities}. While both cities are residentially segregated, they have different segregation \textit{shapes}. In Milwaukee, we see that there is a relatively large concentration of Black residents surrounded by majority White areas. On the other hand, in Atlanta, there is more of a singular boundary that divides the majority Black and majority White areas in half. Do these observations only apply to Milwaukee and Atlanta, or are there common structures of segregation across the United States? This paper uses topological data analysis to explore similar and dissimilar segregation shapes across cities in the United States.

Topological data analysis (TDA) is a form of data analysis that utilizes ideas from algebraic topology to examine the ``shape'' of data. One common way to distinguish two topological spaces is by their number and types of holes. Thus, a circle differs from a disk because a circle has a hole, and a disk does not. Similarly, a coffee cup and a donut are the same ``shape'', as both have one hole. TDA uses these distinctions of topological spaces to study the shape of data and has applications that include studies of biological contagions~\cite{biology}, biological aggregation models \cite{biologicalaggregation, biologicalexperiments}, flow networks in biological transport \cite{biologytransport}, computer vision \cite{computervision}, diurnal cycles in hurricanes \cite{hurricanes}, granular materials \cite{particles}, and many more. Recently, TDA methods have also been utilized in geospatial applications such as studies in voting for the ``Brexit referendum'' \cite{brexit}, global development \cite{globaldevelopment}, city street networks \cite{citystreets}, and voting patterns \cite{levelset}. 

The remainder of this paper is outlined as follows. In Section \ref{background}, we discuss our data set of 2020 U.S. census demographic data, briefly provide an overview of persistent homology, and discuss some related work. In Section \ref{methods}, we outline the methods for this paper, which include two simplicial complex constructions that incorporate the geographic shape of our data. In Section \ref{results}, we explore the results of the clustering of our two simplicial complex constructions. We do additional analyses on mean statistics for our clustering results and compare different clustering methods. Finally, in Section \ref{conclusion}, we explore possible extensions of our work.

\section{Background and Related Work}
\label{background}

\subsection{Demographic Data}\label{sec:demodata}
Our underlying dataset consists of racial and ethnic demographic data for the 112 largest U.S. cities using boundaries defined by the Centers for Disease Control and Prevention \cite{cityboundaries}. We gather census tract shapefiles from the U.S. Census Bureau's TIGER/Line Files \cite{tigerline}. As city boundaries do not align with census tract boundaries, for each city, we consider all census tracts that intersect the region within the city boundary. For our demographic data, we use 2020 Census P2 data from the U.S. Census Bureau \cite{census_data}. 

The P2 table provides ``race alone'' and ``race in combination'' data, including the Hispanic/non-Hispanic ethnicity on the census tract level. The ``race in combination'' category provides counts for people who might mark two or more races on the Census. Thus, if someone identifies as both Asian and White, they are counted both towards the White population count and the Asian population count so people are potentially counted multiple times. On the other hand, the ``race alone'' category only provides counts for people who identify as a single race. While both of these categories have their advantages and disadvantages, in this paper, we only look at ``race alone'' counts, as each value in the data is a singular individual. We further limit our data to only include the racial/ethnic categories of White, Black, Asian, and Hispanic. It is important to note that as Hispanic is an ethnic identity, this category includes people of all races. We choose to include the analysis of Hispanic populations as ethnicity-based segregation is a reality many communities face. While White, Black, Asian, and Hispanic populations are not the only to experience the effects of segregation, they are the central racial/ethnic populations that consistently appear in the cities we choose to study. 

We join the racial/ethnic demographic tabular data to the census tract boundaries of our 112 cities to get a list of shapefiles of each city that contain the census tract boundaries joined with the racial/ethnic demographics for each respective census tract.

\subsection{Persistent Homology} \label{sec:PH}
\textit{Persistent homology} is a computational approach stemming from algebraic topology that seeks to encode the homological features of a data set such as connected components, loops, and trapped volumes \cite{persistenthomology}. Persistent homology (PH) allows one to understand the basic structure of a dataset and thus helps uncover the ``shape'' of data. Broadly, the persistent homology process starts by interpreting a dataset as a noisy sampling of a topological space, iterating over a \emph{filtration} parameter to connect points in the dataset, determining the topological structure made by iterating over the parameter and looking for structures that ``persist" through the parameter filtration. Throughout the filtration, features can be born and die and the difference between a feature's birth and death is the \textit{lifespan} or \textit{persistence}. By looking at the lifespan of features, we often define features with long lifespans to be \textit{signal} in our data and features with short ones to be \textit{noise}.

Suppose we have a dataset $X_\text{observed}$, from which we construct a sequence $X_0 \subseteq X_1 \subseteq \dots \subseteq X_\ell$ of nested topological spaces. 
Each of these topological spaces is an \textit{abstract simplicial complex} (henceforth referred to as a simplicial complex or just a complex), a collection $\mathcal{K}$ of nonempty finite sets, subject to the condition: if $\sigma \in \mathcal{K}$, then every subset of $\sigma$ is also in $\mathcal{K}$. We require that the sequence $\{ X_i \}$ is non-decreasing so that it forms a \textit{filtered simplicial complex}. We refer to each $X_i$ as a subcomplex. The filtered simplicial complex, along with the inclusion maps between its subcomplexes, is called a \textit{persistence complex} \cite{levelset}. In Section \ref{complexes}, we will delve into two types of complexes and explore their construction in detail.


Within a persistence complex, features are classified based on their \textit{homology group}. A homology group is an algebraic invariant associated with a topological space where the $n$th homology group, denoted $H_n$, of a topological space refers to the $n$th dimensional ``holes'' of the space. Specifically, given a topological space, $H_0$ (zeroth homology) counts the number of connected components, $H_1$ (first homology) counts the number of one-dimensional ``holes'' or loops, $H_2$ (second homology) counts the number of two-dimensional voids, and so on. It is the persistence of these homological features that allows us to understand the ``shape'' of a dataset.

We can represent the PH of a complex as a \textit{persistence barcode} where each feature is represented as an interval with endpoints representing when in the filtration the feature is born and when it dies. Thus, the length of a feature's barcode interval represents the lifespan of the feature; see Figure \ref{fig:levelsetbarcode} for an example of a persistence barcode. Alternatively, we can represent the PH of a complex equivalently as a \textit{persistence diagram}, a multi-set of points in the plane where the $x$-value is the birth time of a given feature and the $y$-value is the death time; see the left-most image of Figure \ref{fig:piconversionpipeline} to see an example of a persistence diagram. For foundational material on persistent homology, see \cite{persistenthomology, edelsbrunner2010persistent, edelhare, barcodes}.

As we want to use persistent homology to understand the ``shape'' of segregation, we need to give a basis for interpreting PH results for different homology groups. As our data is two-dimensional, we only consider the zeroth and first homology groups. Thus, consider a complex constructed for a city $X$ and a race $Y$. Then, $H_0$ counts the number of racial enclaves in the city, and $H_1$ counts the number of holes within those enclaves. Persisting $H_0$ features are considered significant enclaves for race $Y$ and are potentially areas of prominent segregation. Similarly, persisting $H_1$ features are large holes where there is not a prominent $Y$ population. Thus, a persisting $H_1$ feature is considered to be an area where non-$Y$ populations are concentrated. These two interpretations of the $H_0$ and $H_1$ groups give us a basis for interpreting persistent homology results in the context of racial segregation.

\subsection{Related Work} \label{sec:relatedwork}

Mathematical and statistical methods have played a pivotal role in analyzing segregation patterns within and across U.S. cities. For example, Frey and Myers \cite{frey2005racial} offer a thorough examination of segregation trends from 1990 to 2000, utilizing dissimilarity indices to capture these changes. Elbers \cite{Elbers2024} explores the impact of population dynamics on residential segregation, providing insights into the evolving patterns of urban segregation. Beyond pattern analysis, several studies have focused on quantifying segregation. Notably, Massey and Denton \cite{dimensionsofsegregation} present a framework consisting of five dimensions to measure residential segregation in the United States, which deepens our understanding of its complexities. Sousa and Nicosia \cite{Sousa2022} utilize statistics from random walks on graphs related to spatial systems to quantify ethnic segregation in cities. Additionally, {\"O}sth, Clark, and Malmberg \cite{sth2014} employ $K$-Nearest Neighbor aggregates to investigate how the scale of administrative boundaries affects segregation outcomes.

Topological data analysis has recently found novel applications in the analysis of geospatial data. In \cite{brexit}, persistent homology was employed to uncover voting patterns in the 2016 European Union ``Brexit'' referendum in the United Kingdom. Similarly, Banman and Ziegelmeier leveraged persistent homology in \cite{globaldevelopment} to examine the correlation between geographic location and national development using global development data. Moreover, Duchin, Needham, and Weighill applied persistent homology in \cite{gerrymandering} to assess redistricting plans in Pennsylvania and North Carolina, identifying features that indicate partisan advantage. While previous studies have applied persistent homology to point cloud data derived from geospatial sources, recent research has started to directly harness the inherent spatial characteristics of geospatial data. For instance, in \cite{citystreets}, Feng and Porter introduce topological methods specifically designed for spatial systems, demonstrating their application to diverse contexts such as city street networks, snowflake structures, and spider webs created under the influence of various psychotropic substances. Moreover, they provide an in-depth analysis of their methods by applying them to California precinct-level voting data from the 2016 presidential election in \cite{levelset}.

In work more closely related to our own, Kauba and Weighill in~\cite{us_demographics} use TDA to study the shape of Black and Hispanic populations in the 100 largest U.S. cities to measure and describe demographic patterns and changes across ten years. In their paper, Kauba and Weighill perform $H_0$ persistent homology on dual graphs constructed from the census tracts of a city and cluster the cities based on the Wasserstein distance between persistence diagrams. For both Black and Hispanic populations, Kauba and Weighill found that cities were clustered with the following categorizations: outliers with many connected components, cities with few connected components, cities with one connected component, and cities with low Black/Hispanic populations. Nevertheless, we propose our independently developed procedure that includes two methods to construct complexes, multiple racial groups, and implements clustering based on a vectorization of persistent homology.

\section{Methods}
\label{methods}

Our overall methodology consists of the following procedure: (1) gathering and preprocessing the necessary data, (2) constructing the complexes used for computing persistence, (3) creating a vector representation, called a persistence image, of the topological summary computed in the second step for $H_0$ (connected components) and $H_1$ (1-dimensional holes), and (4) clustering the cities based off of the vectorized PIs. Steps (1) through (3) are done for each city-race pair, meaning that for 112 cities and four racial groups, we end up with a total of 896 PIs (for each city-race pair, we have separate PIs for $H_0$ and $H_1$). Thus, each city has eight corresponding PIs, and in step (4), we vectorize each of the 8 PIs and concatenate them. Thus, we represent each city as a single vector for clustering purposes.

In this section, we discuss each of the steps introduced above in more detail. We split this section into three subsections: Constructing Complexes, Persistence Images, and Clustering, each corresponding to a respective step in the procedure outlined above (we omit the discussion of data preprocessing here as it was covered in Section \ref{sec:demodata}). Along with explaining the steps generally, we will use a recurring example of the Chicago-White city-race pair to illustrate the methodology more concretely. The GitHub repository containing all scripts and tools used in this study can be found at \cite{friesen_github}.

\subsection{Constructing Complexes}
\label{complexes}

As discussed in Section \ref{sec:PH}, persistent homology can be computed on a nested sequence of topological spaces. While many construction methods utilize the structure of point cloud data, as our data takes the form of census tract polygons, we cannot directly use these methods. To compute persistent homology on spatial data, we could reduce each census tract to a point and construct simplicial complexes from those points. However, doing so would mean we lose important geospatial information about the shapes of our census tracts. Alternatively, we could preserve the geospatial aspect of our data by constructing complexes that do not reduce polygons into points and instead exploit the shape of our data. Below, we provide two complex construction methods that preserve the shape of our data that we will use in this paper: level-set complexes and cubical complexes.

\subsubsection{Level-Set Complexes}

\begin{figure}[htbp]
    \centering
    \begin{minipage}{0.3\textwidth}
        \centering
        \fbox{\includegraphics[width=\linewidth]{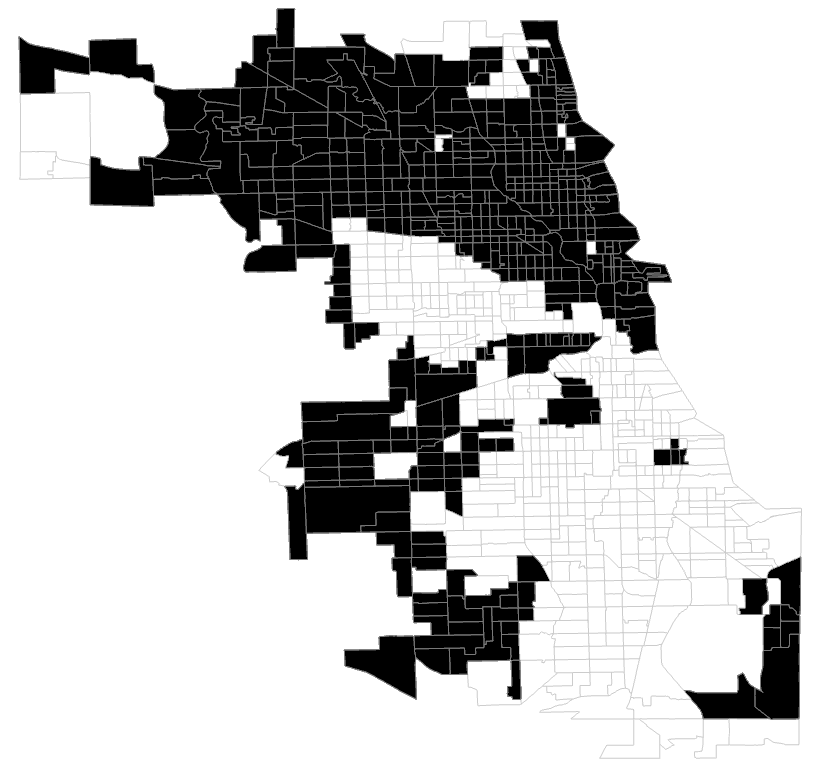}}
    \end{minipage}
    \begin{minipage}{0.1\textwidth}
        \centering
        \raisebox{0.5\height}{\tikz[baseline]\draw[ultra thick,->] (0,0) -- ++ (1,0);}
    \end{minipage}
    \begin{minipage}{0.3\textwidth}
        \centering
        \fbox{\includegraphics[width=\linewidth]{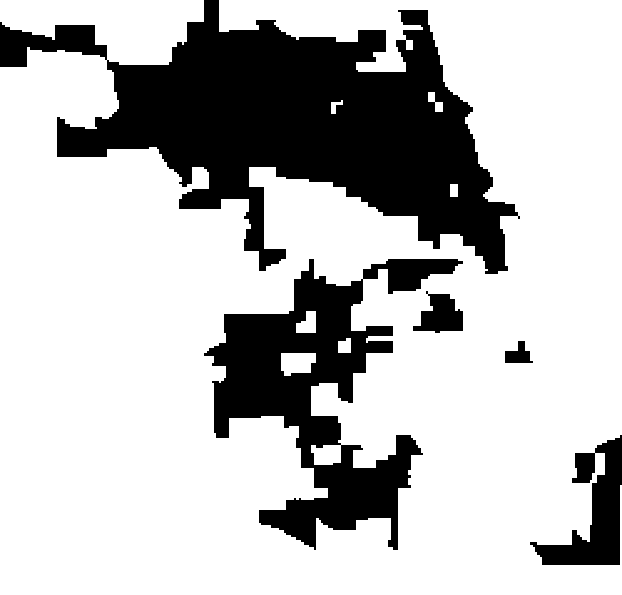}}
    \end{minipage}
    \caption{The procedure for transforming the city's shapefile  (left) into a TIFF image (right) for the Chicago-White pair. The shapefile displays census tract boundaries, with those shaded in black representing areas with a majority White population, while the white-colored tracts have a different demographic majority. For the TIFF image, we convert and include only the tracts that are colored black in the shapefile.}
    \label{fig:shapefiletotiff}
\end{figure}

\textit{Level-set complexes} are based on level sets that directly leverage the geospatial data's manifold nature. We use a given city's shapefile as input to the level-set evolution method. The level-set method for expanding surfaces was first introduced in~\cite{levelsetmethod} but was used to construct complexes by Feng and Porter in~\cite{levelset}. In their paper, Feng and Porter develop level-set complexes to examine ``voting islands'' in 2016 election voting maps of California counties. We give an overview of the level-set method and how we use it in our analyses in this subsection.

Let $M$ denote a 2-dimensional manifold. In our analyses, this manifold consists of the census tracts of a city that has a majority population of a specific race as defined by the city-race pair. For instance, if we are looking at the Chicago-White pair, our manifold would be the census tracts of Chicago that have a majority White population. We construct a sequence of manifolds
\begin{align*}
    M = M_0 \subseteq M_1 \subseteq M_2 \subseteq \dots \subseteq M_\ell 
\end{align*}
by examining the boundary $\Gamma$ of $M$ and performing front-propagation on it so that the boundary of the initial $M$ expands outwards. The level-set method for front-propagation consists of evolving a function $\phi(\vec{x}, t): \mathbb{R}^2 \times \mathbb{R} \to \mathbb{R}$ according to the level set equation
\begin{equation}
    \frac{\partial \phi}{\partial t} = v \left\lvert \nabla \phi \right\lvert
\end{equation}
where $v$ is a velocity constant, $t$ is a time-step parameter, and $\vec{x}$ is the vector normal to the curve. Essentially, the level-set method expands the boundary $\Gamma$ of the manifold $M$ with parameter $t$. We iterate until time $T$ such that $M_0$ is our initial manifold and $M_T$ is our final manifold after front-propagation.

\begin{figure}[htbp]
    \centering
    \includegraphics[width=0.8\linewidth]{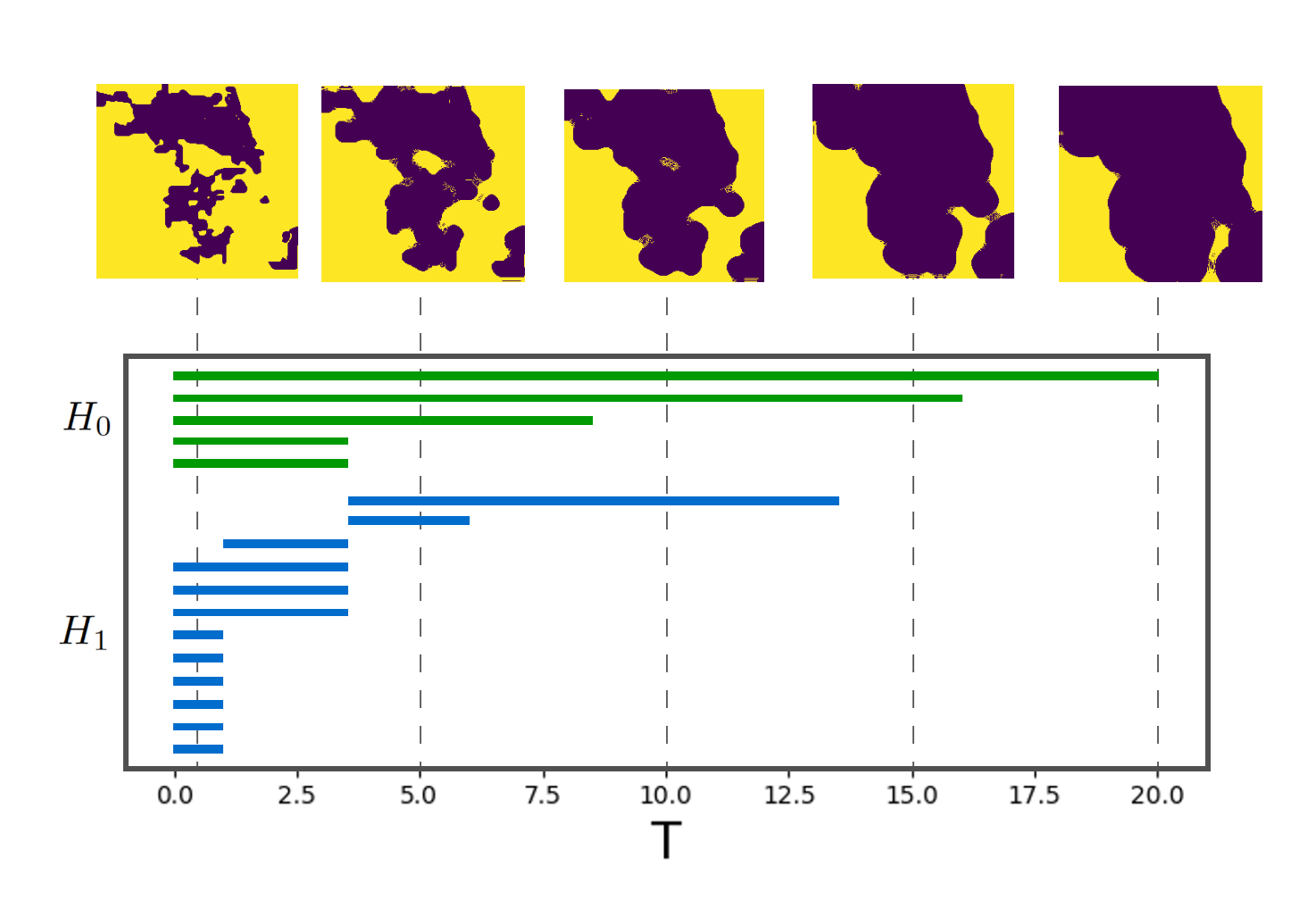}
    \caption{\label{fig:levelsetbarcode} An example of developing a persistence barcode from the level-set boundary expansion of a manifold using the Chicago-White census tracts. The images above the barcode show how the purple manifolds and the yellow empty space develop throughout the level-set process. We then find the persistence of the connected components ($H_0$ features) and topological loops ($H_1$ features) and represent it as a persistence barcode.}
\end{figure}

We will now outline how we implement the level-set method for creating level-set complexes. First, given an $X$-$Y$ city-race pair, we filter the city census tracts $X$ to include only those with a majority $Y$ population. 
Then, we convert this filtered shapefile to a binary raster image and perform our level-set boundary expansion with a velocity of $v=1$ for timesteps $t \in 0,0.5,1,1.5,\dots,20$. To perform the boundary expansion, first, we compute numerical derivatives of our initial image using the Godunov scheme (see~\cite{godunov}). This gives us gradient magnitudes that represent the rate of changes of the level-set function. We then use the gradient magnitudes to compute the change in the level-set function for the time step. We add this change to our original image to get the resulting boundary-expanded image. We repeat this process for all $t$. In Figure \ref{fig:shapefiletotiff}, we show the conversion for the Chicago-White pair from the shapefile to the TIFF image. Once we do this conversion, we perform the level-set boundary expansion shown in Figure \ref{fig:levelsetbarcode}.

Once we perform this level-set expansion, we use the images to build our filtration for computing persistent homology. As seen in Figure \ref{fig:levelsetbarcode}, the manifolds $M_0, M_{0.5},M_1,\dots,M_{20}$ are a nested sequence of topological spaces as $M_0 \subseteq M_{0.5} \subseteq M_1 \subseteq \dots \subseteq M_{20}$. Following the procedure in \cite{levelset}, we convert the level-set expansion images into a filtered simplicial complex by triangulating the images such that every fifth pixel within $M_t$ is a vertex, and each vertex is connected to its cardinal, northwest, and southwest neighbors if the neighbor is also within the boundary expansion at time $t$. Persistent homology is then applied to the resulting clique complex as we iterate over the timestep parameter. Note, with level-set complexes, all connected components will be born at $t=0$ and merge (die) throughout the filtration. In contrast, topological loops can be born and die throughout the filtration. Figure \ref{fig:levelsetbarcode} gives an example of a barcode representation for computing persistent homology using this level-set complex construction. We implemented our methods in Python by adapting the code from \cite{mhcfeng_precinct_2024} to suit our specific applications.

\subsubsection{Cubical Complexes}
The second type of filtration we construct is a nested sequence of \textit{cubical complexes}. Cubical complexes are a collection of elements consisting of points, edges, squares, cubes, and so on which can be written as a union of elementary cubes. This allows us to create elements from data in a grid-like structure. Typically, this grid-like data is in the form of a 2-dimensional or 3-dimensional image where each pixel consists of cubical elements. We generate a nested sequence of cubical complexes as follows: (1) a vertex $v$ is included if $u(v) \geq \varepsilon$ for some function $u$, (2) an edge is included at scale $\varepsilon$ if both of its incident vertices are included, and (3) a square is included at scale $\varepsilon$ if all four of its vertices are included. Persistent homology is then computed from the superlevel set filtration of this cubical complex. For more information on cubical complexes, see \cite{cubical_complex_axioms, cubicalthesis}.

With the level-set complexes discussed above, we require a criterion for selecting which census tracts to include in the boundary expansion. In the case of this paper, for a given $X$-$Y$ city-race pair, we chose the census tracts of city $X$ that have at least $50\%$ population of race $Y$. Thus, changing our selection criterion would affect the results of our level-set method. With cubical complexes, we can instead iterate over the specific criterion boundary to form our filtration. To do this, we iterate over $\varepsilon$ from 100 to 0 in increments of 1. For each $\varepsilon$ in this list, we select all census tracts in city $X$ that have a $Y$ population of at least $\varepsilon$ percent. 
This creates a grayscale image where the color of each census tract represents the percentage of people who identify as race $Y$. Once we produce a grayscale image for the $X$-$Y$ city-race pair, we perform persistent homology from the filtered cubical complex represented as the grayscale image. In our analysis, we utilize the cubical persistence functionality provided by the \textsc{giotto-tda} library \cite{giotto-tda}.

\begin{figure}[htbp]
    \centering
    \includegraphics[width=0.6\linewidth]{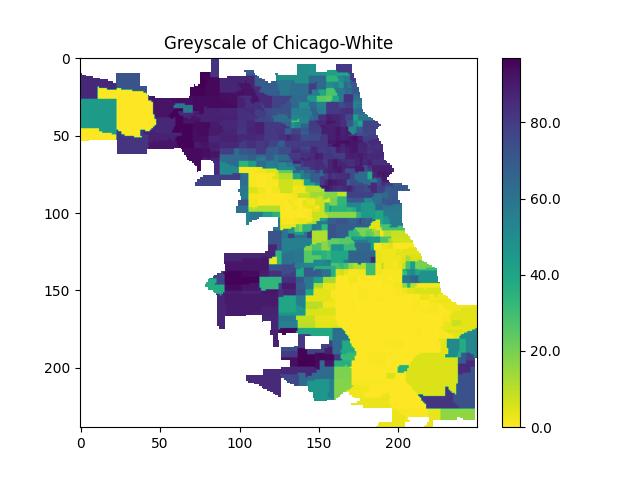}
    \caption{\label{fig:grayscale} The grayscale image of the Chicago-White city-race pair corresponding to the cubical complex filtration. Here, the grayscale image acts as a choropleth map for the White population of Chicago, where yellow census tracts are areas with a lower White population percentage and blue ones represent higher population percentages.}
\end{figure}

\begin{figure}[htbp]
    \centering
    \includegraphics[width=0.8\linewidth]{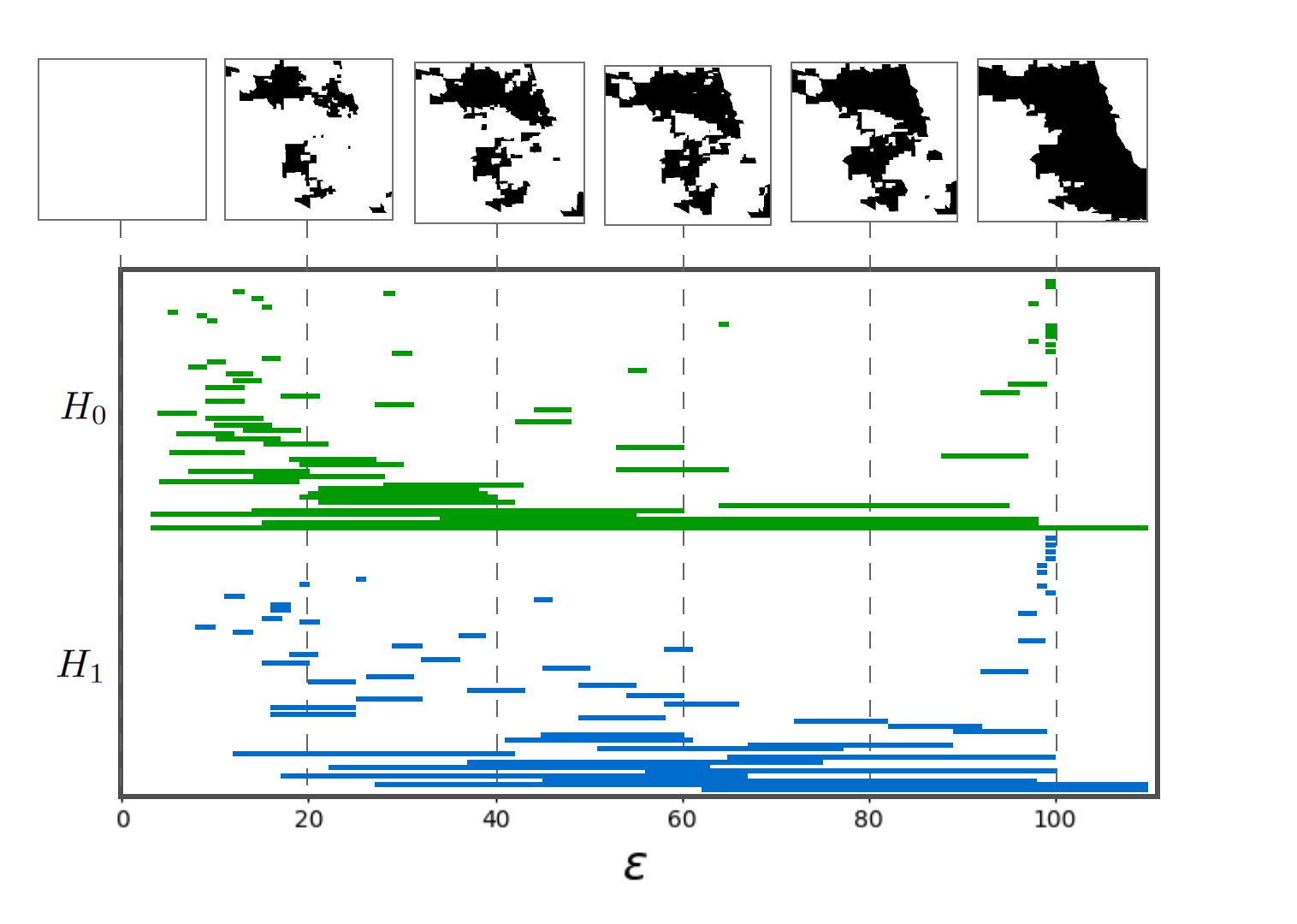}
    \caption{\label{fig:cubicalbarcode} An example of developing persistence through cubical complexes for the Chicago-White city-race pair. The images above the barcode are the census tracts corresponding to the respective $\varepsilon$ value. Compared to the barcode in Figure \ref{fig:levelsetbarcode}, the cubical persistence barcode contains many more features. Additionally, while $H_0$ features were only born at $t=0$ for level-set persistence, with cubical persistence, $H_0$ features can be born at any $\varepsilon$.}
\end{figure}

The grayscale image for the Chicago-White pair can be seen in Figure \ref{fig:grayscale}, and the resulting persistence barcode can be seen in Figure \ref{fig:cubicalbarcode}. One thing to note is that the persistence of a city-race pair using a cubical complex filtration has significantly more features than that of the level-set complex filtration. This is because the level-set filtration eventually loses the structure of the census tracts of a city. In contrast, the cubical complex method maintains the structure and all the intricacies of a city's municipal boundary. Additionally, in Figure~\ref{fig:cubicalbarcode}, we see that there are holes that persist infinitely. This can also be attributed to the geography or intricacies of city boundaries; as with Chicago, two census tracts are wholly contained within the city limits of Chicago but are not officially part of Chicago.

\subsection{Persistence Images}
Once we have computed the persistent homology for all 112 of our cities, we need to utilize the information from the persistence as input to our clustering algorithms. A persistence diagram is a multi-set of points in the plane, and thus not amenable to many machine learning methods. There exist notions of distance for comparing persistence diagrams such as the bottleneck or Wasserstein, but they are very computationally intensive \cite{dey2022computational}. As such, a number of topological summaries, have been introduced to transform a persistence diagram into an object in a vector space. These include persistence landscapes~\cite{bubenik2015statistical}, persistence scale-space kernel~ \cite{reininghaus2015stable}, persistence weighted Gaussian kernel~\cite{JMLR:v18:17-317}, sliced Wasserstein kernel~\cite{carriere2015stable}, persistence Fisher kernel~\cite{DBLP:conf/nips/LeY18}, and persistence images~\cite{persistenceimages}.

We use persistence images (PIs)~\cite{persistenceimages} in our analysis as they are a transformation of a persistence diagram into a vector in Euclidean space which have a highly interpretable connection to the persistence diagram and are stable. This means that they provide a way to represent the persistent homology of a dataset as a vector which remain relatively unchanged to small perturbations in the input. To create a persistence image from a persistence diagram, we follow the PI pipeline using the \textsc{Persim} Python package \cite{persim}:
\begin{enumerate}
    \item Convert the persistence diagram from birth-death coordinates to birth-persistence coordinates with the linear transformation $\tau: \mathbb{R}^2 \to \mathbb{R}^2$ by $\tau(x,y) = (x,y-x)$. For each coordinate in our PD, $\tau$ maintains the birth and defines the new $y$-coordinate to be the difference between birth and death, or in other words, the lifespan.
    \item For each point $u = (u_x, u_y)$ in our birth-persistence PD, we center a normalized Gaussian distribution, 
    \begin{align*}
        g_u(x,y) = \frac{1}{2\pi \sigma^2}e^{\frac{-\left( (x-u_x)^2 + (y-u_y)^2 \right)}{2\sigma^2}}
    \end{align*}
    to create a persistence surface. We use $\sigma = 1$ for level-set PIs and $\sigma = 5$ for cubical PIs. We choose different values of $\sigma$ to ensure consistent relative variance between the level-set and cubical PIs, which have different resolutions. 
    \item We then utilize a weighting function to ensure stability of the PIs that is continuous, piecewise differentiable, and zero for points of zero lifetime. We choose to weight points of larger lifespan higher, as high-persistence features are most relevant to the overall shape. To do this, we use a standard linear ramp weight function. Thus, our persistence surface is weighted such that features with higher persistence have a larger impact on our image.
    \item Next, we overlay a grid of a chosen resolution onto our persistence surface of the weighted normalized Gaussian distributions. For our purposes, we choose a resolution of $20 \times 20$ when using level-set complexes and $100 \times 100$ when using cubical complexes. We choose these resolutions as they align with the time steps used in each complex construction.
    \item For each pixel in our grid, we take the integral of our persistence surface to get the resulting grayscale pixel values for our image.
\end{enumerate}

\begin{figure}[htbp]
    \centering
    \includegraphics[width=0.9\linewidth]{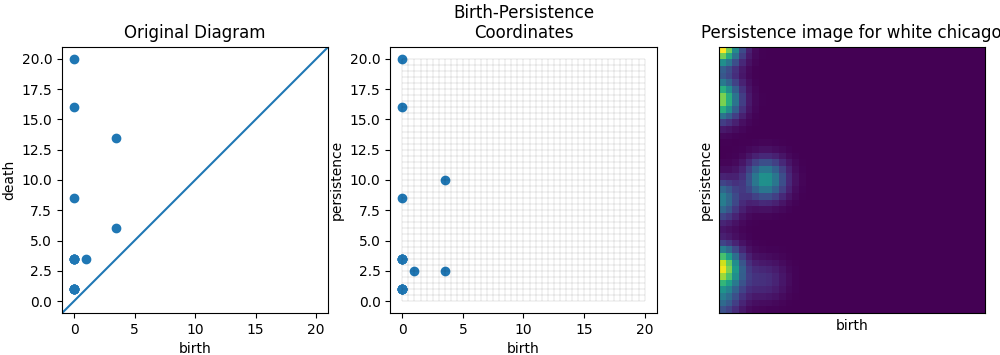}
    \caption{\label{fig:piconversionpipeline} An example of the pipeline of converting a persistence diagram to a persistence image for the Chicago-White level-set $H_1$ homology. We start with the original diagram (left) and perform a linear transformation to convert the diagram from birth-death coordinates to birth-persistence coordinates (center). We then convert the birth-persistence diagram to a persistence image (right) by centering a normalized Gaussian at each point in our persistence diagram.}
\end{figure}

As it is impossible to distinguish $H_0$ from $H_1$ features in a PI, we must create a distinct PI for the $H_0$ and $H_1$ homology classes. Thus, for the 112 cities and four race groups, we have 896 PIs (448 $H_0$ PIs and 448 $H_1$ PIs). Additionally, as there is no way to represent features that die at infinity in a finite representation, we limit these infinite deaths to 20 for level-set persistence and 100 for cubical persistence. Figure \ref{fig:piconversionpipeline} shows how we convert the Chicago-White level-set $H_0$ persistence diagram to a persistence image. Given a collection of PIs of resolution $N \times N$, we can vectorize the images to get a collection of vectors of dimension $N^2$. As these images are now vectors in $\mathbb{R}^{N^2}$, we can use them as input for various machine-learning techniques.

\subsection{Clustering}
    \begin{figure}[htbp]
    \centering
    \begin{subfigure}{.5\textwidth}
      \centering
      \includegraphics[width=1\linewidth]{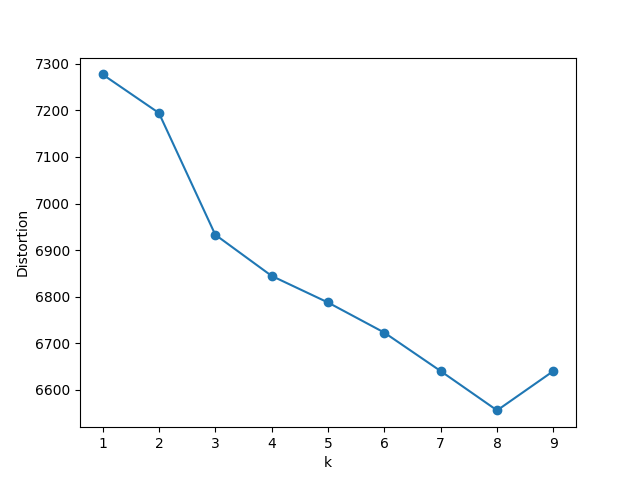}
      \caption{Level-Set Complex}
    \end{subfigure}%
    \begin{subfigure}{.5\textwidth}
      \centering
      \includegraphics[width=1\linewidth]{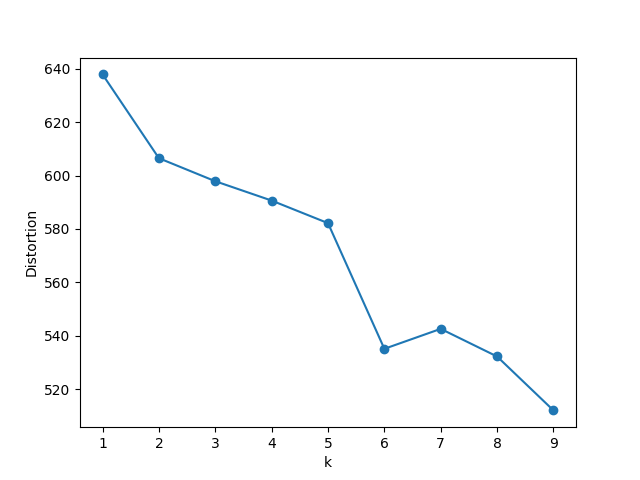}
      \caption{Cubical complex}
    \end{subfigure}
    \caption{Elbow plots for $K$-medoids clustering of cities based on PIs for the given complex construction methods.}
    \label{fig:elbowplots}
\end{figure}

Once we have PIs for each city-race pair for our two complex constructing methods, we perform clustering to determine which cities have similar segregation shapes. For our purposes, we choose to classify cities using \textit{$K$-medoids clustering} \cite{kmedoids} rather than \textit{$K$-means clustering}. In $K$-means clustering, each datapoint is assigned to a different \textit{centroid}, where the coordinate of each centroid is the mean of the coordinates of the objects within the cluster. The $K$-means algorithm iterates over different centroids to find the $K$ centroids that minimize the sum of squares of dissimilarities for each cluster. While $K$-means is computationally efficient, it is known to be sensitive to the existence of data outliers. Thus, we use $K$-medoids clustering, where clusters are represented by their \textit{medoids} rather than centroids. A cluster medoid is an object of the dataset that is the most centrally located among the objects within the cluster. Along with using medoids rather than centroids, $K$-medoids clustering seeks to minimize average dissimilarities instead of sums of squares of dissimilarities \cite{kmedoidsalgorithm}. For all computations related to $K$-medoids clustering and its analyses, we use the Python library \textsc{scikit-learn} \cite{scikit-learn}.

For each city, we have 8 PIs ($H_0$ and $H_1$ PIs for each racial group), which means that there are many ways we can determine how we cluster our cities. For example, we can include all the information by concatenating all eight vectorized PIs into one vector for each city. Doing this will give us clustering results considering the segregation structure for all four racial groups. However, perhaps we only want to look at the structure of White and Black segregation as this is what has been the most prominent form of segregation through redlining~\cite{massey1993american}. Thus, we might only want to concatenate the White and Black PIs as our input for clustering. Many combinations can be used, and they all serve different purposes. Our paper focuses on the following concatenated groups: White, Black, Asian, and Hispanic; White and Black; and Black, which we will denote as WBAH, WB, and B, respectively. We concatenate the vectorized $H_0$ and $H_1$ PIs for each of these groups and use this new vector as the input for our clustering methods. 

We utilize a common procedure for determining values of $K$ called \textit{elbow plots} \cite{bishop}. For a range of values of $K$, we compute the sum of distances from points to their medoids, known as the distortion. We then choose a value of $K$ at a point where the decrease in distortion starts to slow down, known as the ``elbow''. Figure \ref{fig:elbowplots} shows the elbow plots for level-set and cubical clustering. We see that for level-set clustering, after 3 clusters, the change in distortion is seemingly linear, indicating that $K \geq 3$ is sufficient. For cubical clustering, distortion is fairly linear for $K$ between 2 and 5, but there is a significant drop from $K=5$ to $K=6$. While this drop indicates that $K=6$ would be ideal for cubical clustering, we use $K=4$ to prioritize ease of interpretation. This choice provides a balance between sufficient granularity and manageable complexity, allowing for meaningful insights without overwhelming detail. 

\section{Results and Analysis}
\label{results}
Now that we have detailed our methodology, we describe the results of our study and perform some analysis of our results. We split this section into four subsections. The first two explore the results from using level-set and cubical complexes as inputs to our clustering algorithm. In the third subsection, we look at mean demographic statistics for the clusters as found in the first two subsections. This gives us a basic understanding of whether a correlation exists between our methodology's results and some of the underlying data used in our methodology. We then seek to compare the resulting clusters between different input considerations in the final section.

\begin{figure}[htbp]
    \centering
    \includegraphics[width=0.8\linewidth]{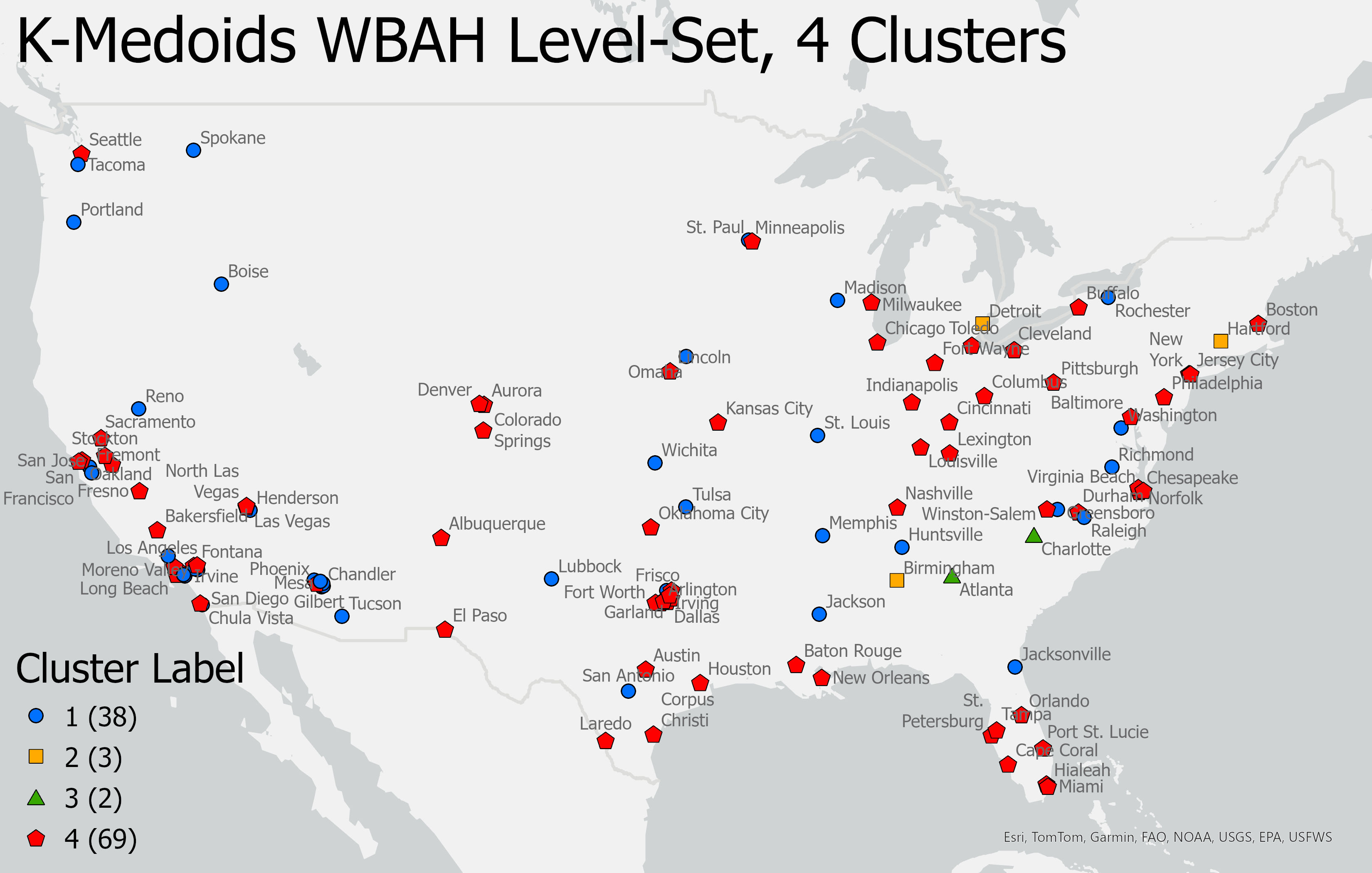}
    \caption{\label{fig:wbahlevelsetmap} A map showing the results of the $K$-medoids clustering of WBAH level-set persistence. Each of our 112 cities is labeled based on the cluster they were assigned by the $K$-medoids algorithm. The legend in the bottom left shows the cluster numbers and the number of cities in the cluster.}
\end{figure}

\begin{figure}[htbp]
    \centering
    \begin{tabular}{ll}
    \includegraphics[width=.65\linewidth,valign=m]{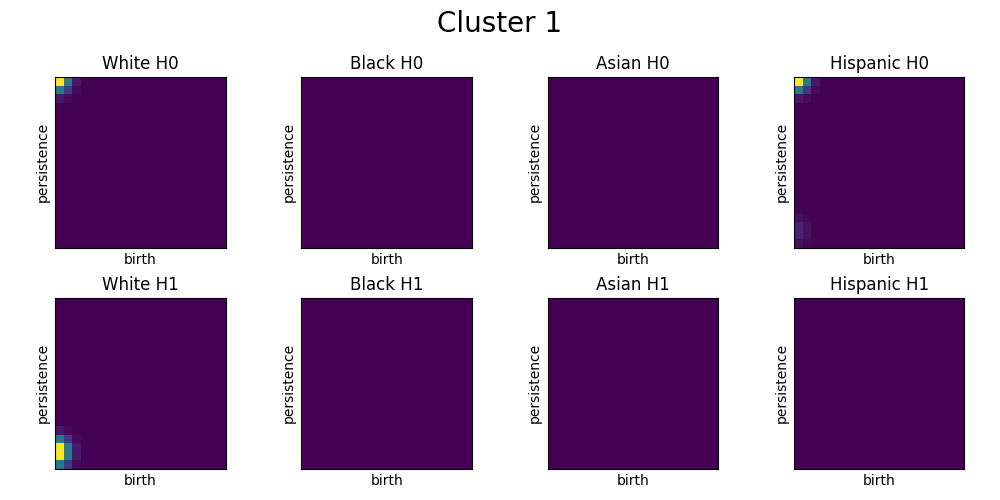} \\\includegraphics[width=.65\linewidth,valign=m]{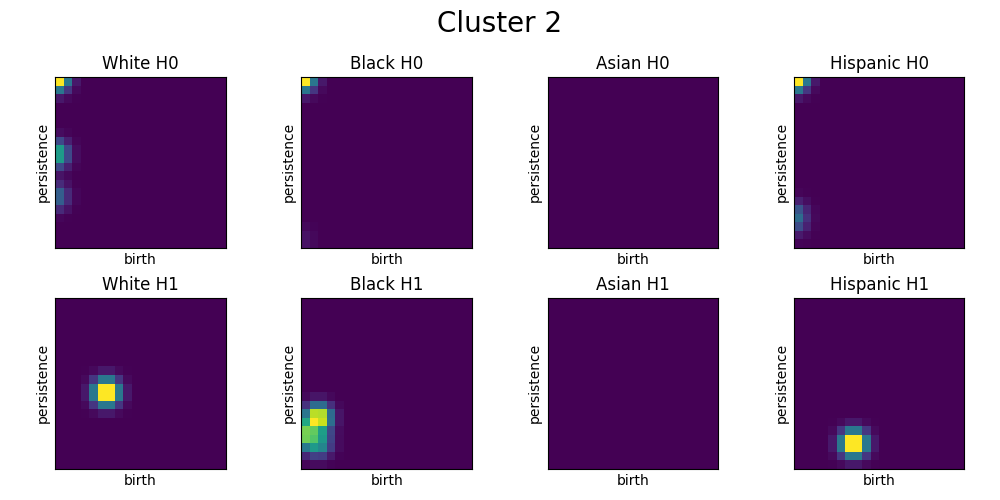}\\

    \includegraphics[width=.65\linewidth,valign=m]{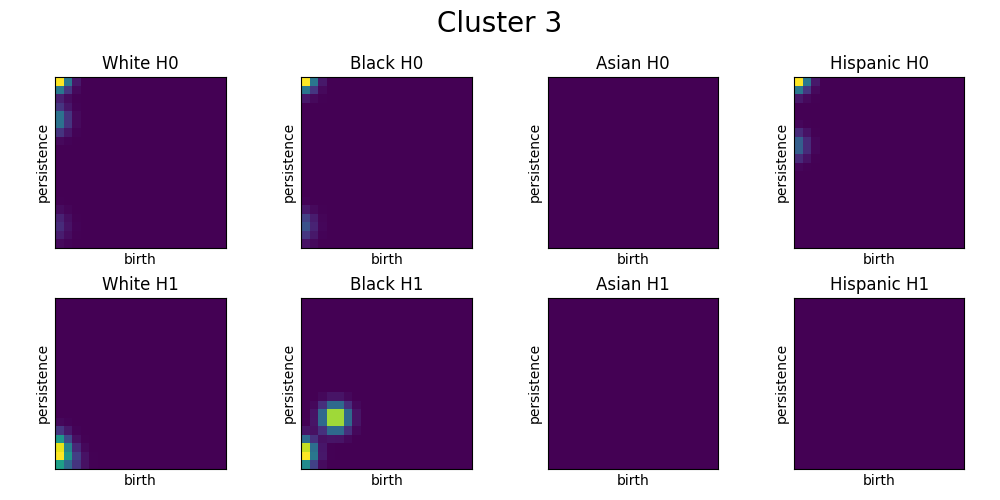} \\\includegraphics[width=.65\linewidth,valign=m]{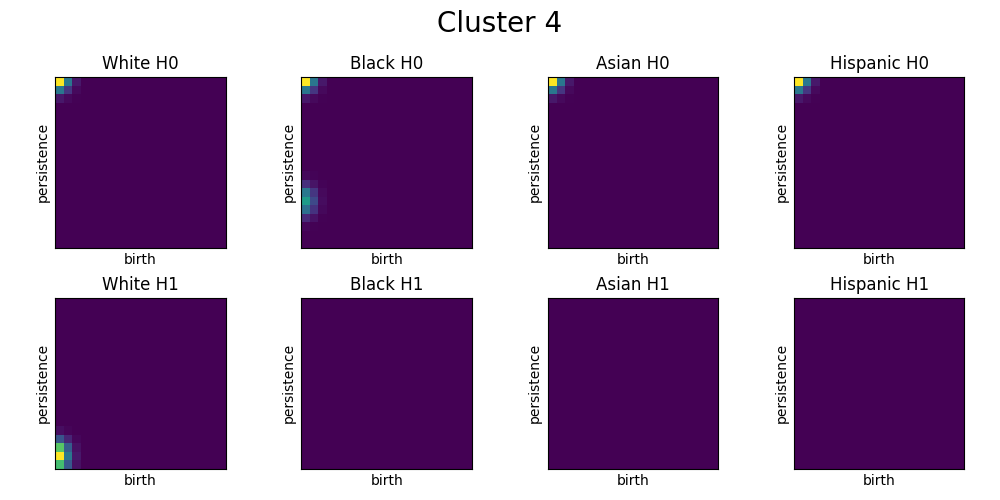}\\
    \end{tabular}
    \caption{\label{fig:ls_wbah_4_centroids} A list of the PIs  corresponding to each $K$-medoids cluster for WBAH level-set homology. Each set of 8 PIs is the PIs of the medoid city for each cluster where the cluster number is the same as in Figure \ref{fig:wbahlevelsetmap}. Each cluster has a PI for $H_0$ and $H_1$ homology for each racial group.}
\end{figure}

\subsection{Level-Set Complex Clustering Results}
We first use WBAH (White, Black, Asian, Hispanic) level-set persistence as input to $K$-medoids clustering using 4 clusters. Figure \ref{fig:wbahlevelsetmap} shows a map of the resulting clusters. We see that most cities are a part of Cluster 4, and Clusters 2 and 3 only contain 3 and 2 cities, respectively. Using $K$-medoids clustering, we determine which cities are the cluster medoids: Henderson for Cluster 1, Hartford for Cluster 2, Atlanta for Cluster 3, and Lexington for Cluster 4. While we can see the results in Figure \ref{fig:wbahlevelsetmap}, this map cannot tell us much about why the cities are clustered the way they are.

One way we can better understand our clusters is by looking at the PIs of the medoid of each cluster. This provides some insight because the medoid cities are found to be the most representative of the clusters through the $K$-medoids clustering. In Figure \ref{fig:ls_wbah_4_centroids}, we get a sense as to why certain cities are clustered together. For example, the medoid for Cluster 4 differs from that of Cluster 1 because it contains Black and Asian $H_0$ features that persist until $t=20$. This tells us that cities in Cluster 1 generally do not have any census tract with a majority of Black or Asian populations, while those in Cluster 2 do. Since this is the primary distinction between Clusters 1 and 4, we can infer that either cities in Cluster 1 have less segregation because their Black and Asian populations are more dispersed compared to those in Cluster 4, or cities in Cluster 1 simply have smaller Black and Asian populations than those in Cluster 4.

Compared to Clusters 1 and 4, the PIs of the medoids for Clusters 2 and 3 have more complex structures. In both clusters, we see no Asian $H_0$ or $H_1$ features, but features of all other races are present. We see that the difference between Cluster 2 and 3 is the White $H_1$ and Hispanic $H_1$ features in Cluster 2. Here, Cluster 2 has a White hole that appears at around $t=7$ and persists for roughly 10 time steps. Additionally, in Cluster 3, there is no Hispanic hole, while there is one that appears for a little bit in Cluster 2.

As we use $K$-medoids clustering, we must note how we can interpret the medoid PIs. While we can cluster all 112 cities based on these medoids, each city in a given cluster will have different structures. For example, while the Cluster 1 medoid does not have any Black or Asian $H_0$ or $H_1$ features, this does not mean these features are also not present in all cities in Cluster 1. It just means that a given city in Cluster 1 \textit{aligns more closely} with the medoid of Cluster 1 than any other cluster.

\subsection{Cubical Complex Clustering Results}
\begin{figure}[htbp]
    \centering
    \includegraphics[width=0.8\linewidth]{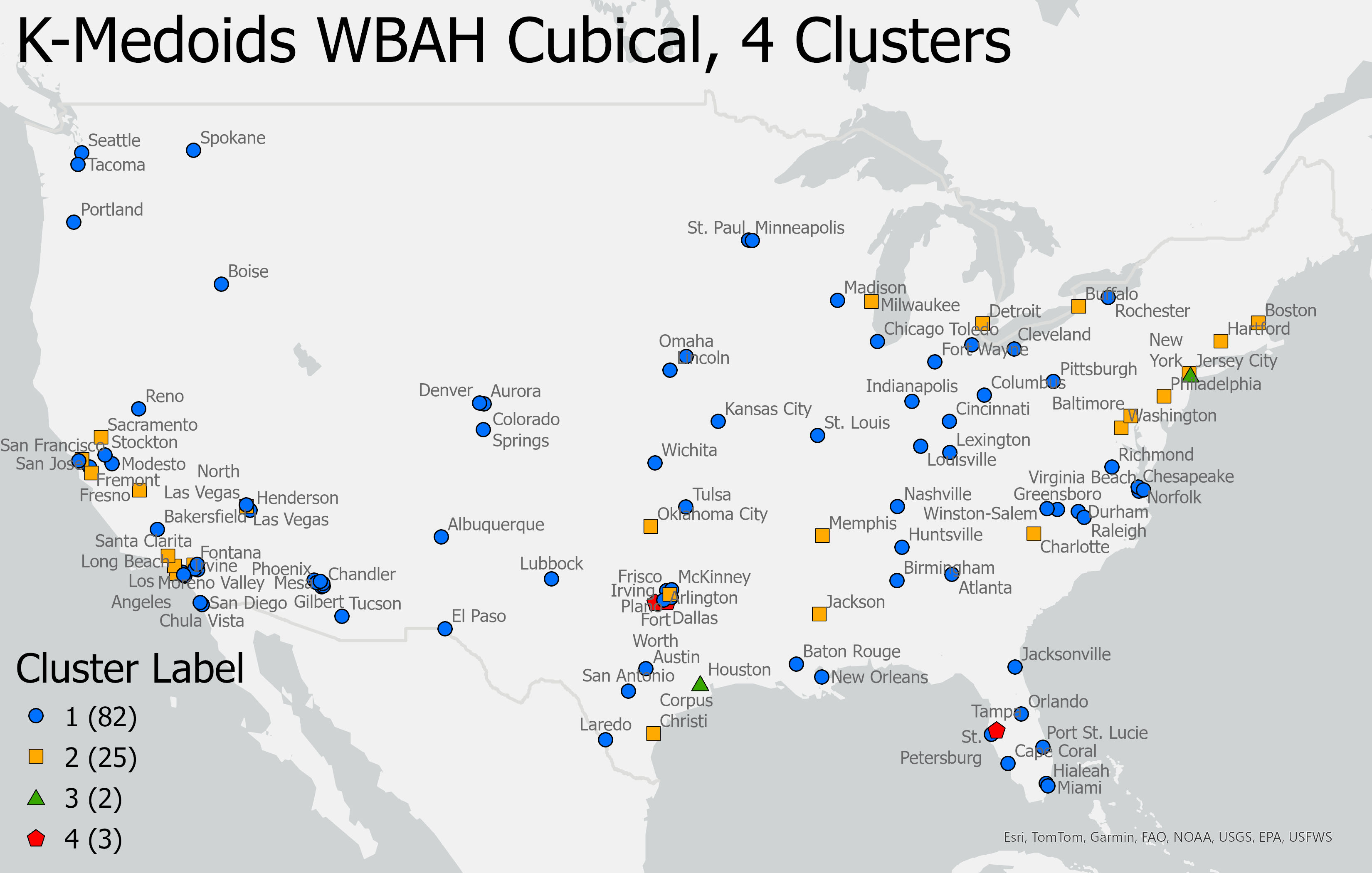}
    \caption{\label{fig:wbahcubicalmap} A map showing the results of the $K$-medoids clustering of WBAH cubical persistence. Each of our 112 cities is labeled based on the cluster they were assigned by the $K$-medoids algorithm. Note that the cluster labels in this map are not related to those from Figure \ref{fig:wbahlevelsetmap}.}
\end{figure}

This subsection uses WBAH cubical persistence as the input to $K$-medoids clustering using 4 clusters. 
Figure \ref{fig:wbahcubicalmap} shows the map of the resulting clusters. Similar to the results of the level-set clustering, we see that most cities are in Clusters 1 and 2. The cluster medoids are Gilbert (Cluster 1), Hartford (Cluster 2), Houston (Cluster 3), and Tampa (Cluster 4). Interestingly, Hartford is selected as a medoid for both the level-set and cubical approaches.

\begin{figure}[htbp]
    \centering
    \begin{tabular}{ll}
    \includegraphics[width=.65\linewidth,valign=m]{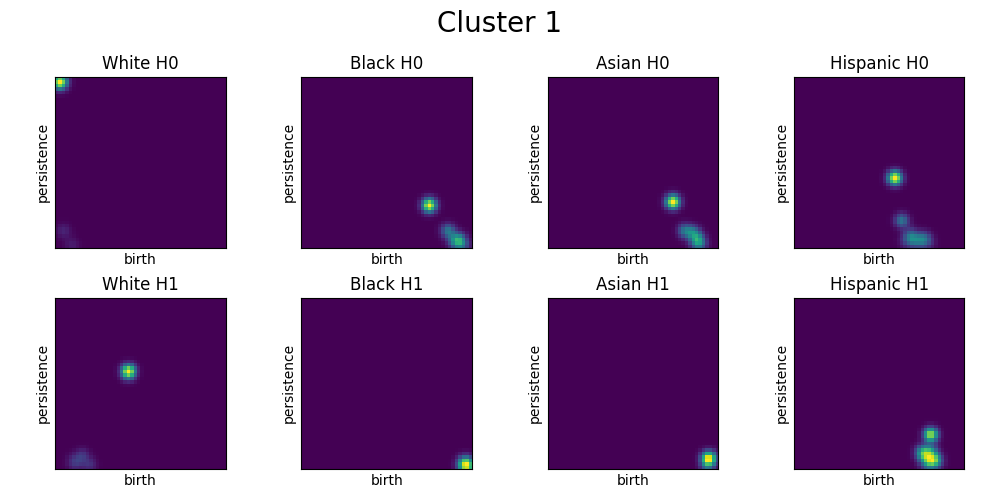} \\ \includegraphics[width=.65\linewidth,valign=m]{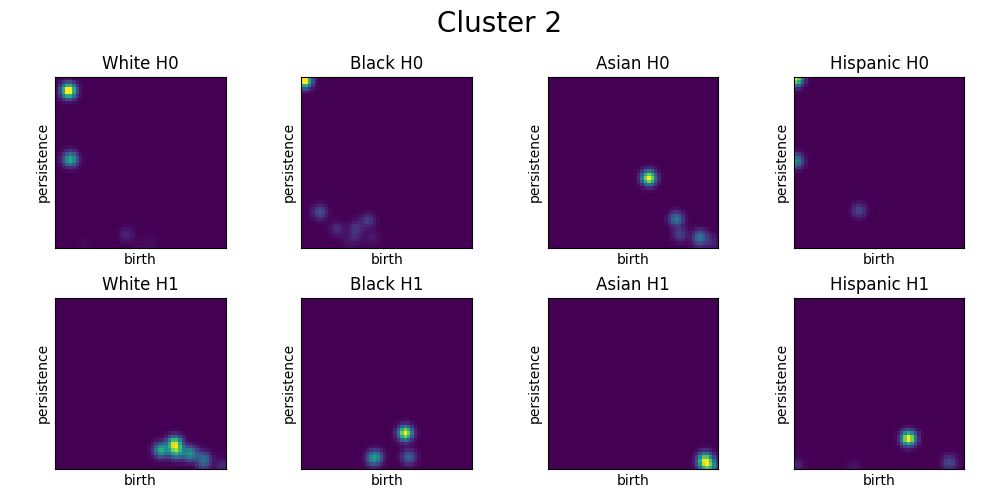}\\

    \includegraphics[width=.65\linewidth,valign=m]{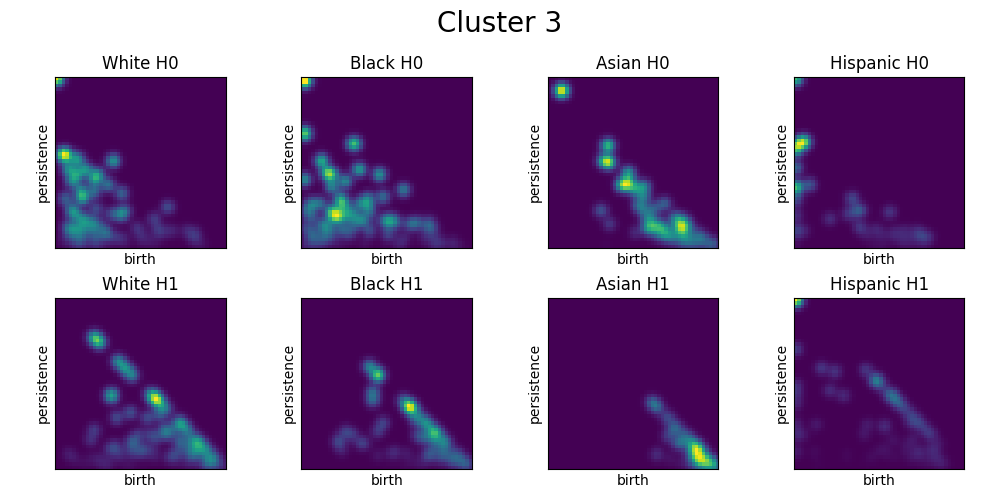} \\ \includegraphics[width=.65\linewidth,valign=m]{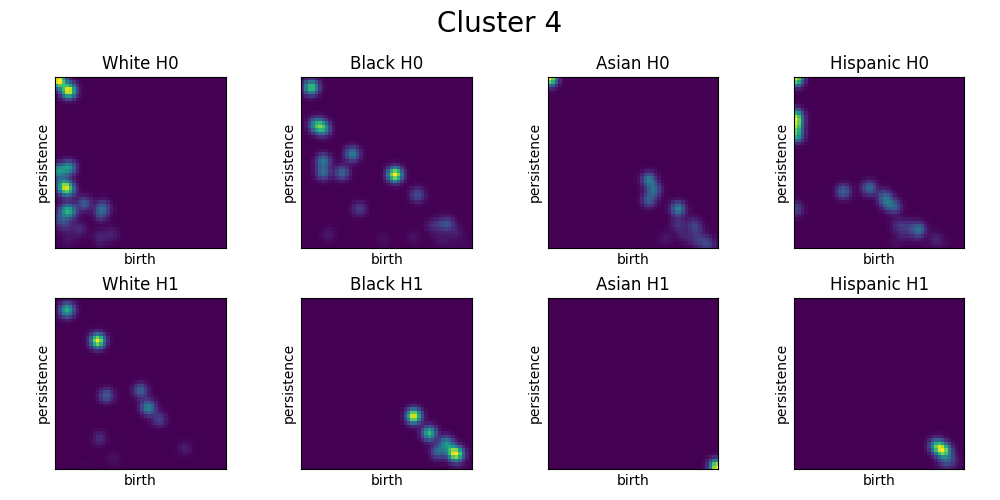}\\
    \end{tabular}
    \caption{\label{fig:cubical_wbah_4_centroids} A list of the PIs corresponding to each $K$-medoids cluster for WBAH cubical homology. Each set of 8 PIs is the PIs of the medoid city for each cluster where the cluster number is the same as in Figure \ref{fig:wbahcubicalmap}.}
\end{figure}

In Figure \ref{fig:cubical_wbah_4_centroids}, we see the medoid PIs for cubical persistence have many more features than the PIs for level-set persistence. First, let us focus on the Cluster 1 and Cluster 2 medoid PIs. We see that the Cluster 1 Black, Asian, and Hispanic $H_0$ PIs look reasonably similar since the first feature of each image appears at higher values of $\varepsilon$. This means that each census tract of the medoid city has a relatively low Black, Asian, and Hispanic population. For Cluster 2, on the other hand, Black and Hispanic $H_0$ features appear in the beginning, meaning that some census tracts have very high Black and Hispanic populations. The presence of both a White $H_0$ feature and a Black $H_0$ feature with large lifespans for the Cluster 2 medoid implies that Cluster 2 cities have more notable structures of segregation than cities in Cluster 1. Additionally, a persisting White $H_1$ feature in the PIs for Cluster 1 implies that census tracts with higher non-White populations are centralized within the city boundaries and are most likely racial enclaves. As there is no persisting White $H_1$ feature for Cluster 2, we can reason that cities in this cluster are more likely to have a dividing line between White and non-White populations instead of having non-White enclaves within a majority White city.

Clusters 3 and 4, like Clusters 2 and 3 for level-set persistence, are composed of 2 and 3 cities, respectively. Compared to Clusters 1 and 2, the PIs of the medoids in Clusters 3 and 4 exhibit significantly more intricate structures. In Cluster 3, $H_0$ and $H_1$ features are present across all racial groups, with consistent birth of White and Black $H_0$ and $H_1$ features throughout the filtration. While the PI of the Cluster 4 medoid has fewer features than that of Cluster 3, it still shows more complexity than Clusters 1 and 2. Notably, Asian $H_0$ features and Black, Asian, and Hispanic $H_1$ features emerge later in the filtration, whereas the remaining features are born more consistently throughout.

\subsection{Understanding Demographic Statistics of Our Clustering Results}
While observing the PIs of the medoids of our clusters might allow for some qualitative observations, observations stemming from PIs alone are limited. For example, the Cluster 1 medoid for the cubical persistence results has no census tract with a majority Black, Asian, or Hispanic population. We can say the same for the Cluster 1 medoid of the level-set persistence results. Does this imply that cities in Cluster 1 of each method have less segregation because the non-White populations are more dispersed across the census tracts, or do these cities have a low non-White population? We cannot say definitively without further information. 

Along with the demographic data of our cities, many indices exist that quantitatively measure segregation. Two of the most common indices are the Index of Dissimilarity and the Exposure Index (EI). The \textit{Index of Dissimilarity} (IoD)~\cite{segregationindices} is a comparison between groups that measures a group's relative segregation (high dissimilarity) or integration (low dissimilarity) to another group across a city. The IoD formula takes the form
\begin{align*}
    IoD = \frac{1}{2} \sum_{i=1}^n \left\lvert \frac{P_{1,i}}{P_1} - \frac{P_{2,i}}{P_2} \right\rvert
\end{align*}
where $P_x$ is the city-wide population of group $x$, $n$ is the total number of areal units (in this case, census tracts) in the city, and $P_{x,i}$ is the population of group $x$ in area $i$. The IoD is always a value between 0 and 100, which can be interpreted as the percentage of people from either group that would need to move such that the distribution of each group is the same across all census tracts within the city \cite{censusscope_dissimilarity}. On the other hand, the \textit{Exposure Index} (EI)~\cite{segregationindices} measures the average exposure of a member of one group to members of another group. The EI formula takes the form
\begin{align*}
    EI = \sum_{i=1}^n \left( \frac{P_{1,i}}{P_1} \right) \left( \frac{P_{2,i}}{T_i} \right)
\end{align*}
where $P_x$ is the city-wide population of group $x$, $P_{x,i}$ is the census tract $i$ population of group $x$, $T_i$ is the total population of census tract $i$, and $n$ is the number of census tracts in the city. The EI ranges from 0 to 100, where a larger value means that, on average, members of one group live in a census tract with a higher percentage of people from the other group. Thus, a higher EI value implies a higher likelihood of interaction between the two groups, while a low score implies a lower likelihood of interaction. As the Exposure Index takes in two groups, choosing the groups to be the same means the EI measures the likelihood that a member of the group interacts with another member of the group. Thus, low values of the Exposure Index given a single group imply that the group is not isolated from other groups, while high values imply that the group is heavily isolated \cite{censusscope_exposure}.

To better understand the patterns of our clusters, we utilize primary city-wide demographic data along with Index of Dissimilarity and Exposure Index data from \cite{brown_diversity_data} to provide basic summary statistics for our clusters given a specific clustering method. We follow the same techniques used in \cite{geography_statistics} to provide the mean statistics of each cluster from a clustering algorithm. The statistics we include in our cluster analysis are total population, White population percentage, Black population percentage, Asian population percentage, Hispanic population percentage, White-Black IoD, White-Hispanic IoD, White-Asian IoD, White-White EI, Black-Black EI, Hispanic-Hispanic EI, and Asian-Asian EI.

We first convert the raw values for each field listed above to Z-score values. The Z-score is a statistical measurement that gives the distance from an observed value of the dataset to the mean value of the dataset \cite{zscore}. The Z-score is derived from the formula
\begin{align*}
    Z = \frac{x - \mu}{\sigma}
\end{align*}
where $x$ is the observed value, $\mu$ is the mean value of the dataset, and $\sigma$ is the standard deviation of the dataset. The resulting value tells us how many standard deviations the sample is from the mean. A negative value means the sample is below the mean, while a positive value means the sample is above the mean. This gives us a simple way of comparing values across different variables. To analyze the statistics of a given cluster, we calculate the Z-score for each city within the cluster, using the mean and standard deviation derived from all cities. Then, we determine the average Z-score for the cluster by taking the mean Z-score of all its cities.

\def\cca#1{\cellcolor{black!#10}\ifnum #1>5\color{white}\fi{#1}}

Table \ref{table:wbah_levelset} shows the mean cluster statistics for the White, Black, Asian, and Hispanic Level-Set $K$-medoids clustering. Looking at Table \ref{table:wbah_levelset}, the two clusters that stand out are Clusters 2 and 3. We see that Cluster 2 cities have, on average, significantly higher Black and lower White populations than the mean. Additionally, Cluster 2 has high White-Black, White-Hispanic, and White-Asian IoD values, very low White-White EI, and high Black-Black EI. All of this information implies that Cluster 2 cities (Detroit, Birmingham, and Hartford) have high segregation, specifically White-Black segregation. Looking back at the cluster medoid PIs in Figure \ref{fig:ls_wbah_4_centroids}, we see that the long-persisting Black $H_0$ feature aligns with the assumption that there is high White-Black segregation among cities in Cluster 2. Similarly, Cluster 3 cities (Atlanta and Charlotte) have similar mean statistics but have smaller Black populations and lower White-White and Hispanic-Hispanic Exposure Indices.

\begin{table}
    \centering
    \resizebox{\textwidth}{!}{\begin{tabular}{c | R R R R R | R R R | R R R R}
      \bottomrule
       & \multicolumn{5}{|c} {\bf Population} & \multicolumn{3}{|c} {\bf IoD} & \multicolumn{4}{|c} {\bf EI}\\
      \multicolumn{1}{c} {\textbf{Cluster}} & \multicolumn{1}{|c} {Total} & \multicolumn{1}{c} {W} & \multicolumn{1}{c} {B} & \multicolumn{1}{c} {A} & \multicolumn{1}{c} {H} & \multicolumn{1}{|c} {W-B} & \multicolumn{1}{c} {W-H} & \multicolumn{1}{c} {W-A} & \multicolumn{1}{|c} {W-W} & \multicolumn{1}{c} {B-B} & \multicolumn{1}{c} {H-H} & \multicolumn{1}{c} {A-A} \\
      \toprule
      \bottomrule
      \textbf{1} & -0.22 & 0.22 & -0.19 & 0.14 & -0.18 & -0.40 & -0.33 & -0.30 & 0.18 & -0.27 & -0.25 & 0.02\\
      \textbf{2} & -0.27 & -1.98 & 2.36 & -0.71 & -0.33 & 1.12 & 1.05 & 1.08 & -1.00 & 1.85 & 0.17 & 0.08\\
      \textbf{3} & 0.10 & -0.63 & 1.24 & -0.27 & -0.75 & 1.14 & 0.40 & 0.35 & 0.49 & 1.26 & -0.81 & -0.25\\
      \textbf{4} & 0.13 & -0.02 & -0.03 & -0.04 & 0.14 & 0.15 & 0.13 & 0.11 & -0.07 & 0.03 & 0.16 & -0.01\\
      \toprule
    \end{tabular}}
    \caption{\label{table:wbah_levelset} Mean Cluster Statistics for $K$-Medoids White (W), Black (B), Asian (A), and Hispanic (H) Level-Set Clustering. For each cluster, we show the mean population, Index of Dissimilarity, and Exposure Index Z-scores. We choose to only display the IoD between White and the other races as segregation in the U.S. is typically viewed as a divide between White and non-White populations. We choose to include EI between the same racial group as this becomes a measure of isolation. Red cells represent Z-scores below the overall mean, while green cells represent those above the mean.}
\end{table}

In Figure \ref{fig:ls_wbah_4_centroids}, we observed that the only differences between the Cluster 1 and Cluster 4 medoid Persistence Images were that Cluster 4 had persisting Black, Asian, and Hispanic F$H_0$ features while Cluster 1 did not. The results in Table $\ref{table:wbah_levelset}$ show us that there are no significant city-wide demographic differences between Clusters 1 and 4. This means that the differences in the PIs were not solely because of demographic differences; they relate to the differences in the segregation structures. Therefore, our analysis shows that cities in Cluster 1 are more racially integrated than those in Cluster 4. This is further confirmed by the differences in IoD values between the two clusters. Cluster 1 has negative IoD values, meaning Cluster 1 cities are more integrated than the mean. Comparatively, Cluster 4 has slightly higher IoD values than the mean.

We now do a similar analysis for the resulting clusters from the cubical persistence clustering. The mean cluster statistics for this method are shown in Table \ref{table:wbah_cubical}. Recall we first looked at the cubical clustering results in Figure \ref{fig:wbahcubicalmap}. We noted that Clusters 1 and 2 accounted for most of the cities while Clusters 3 and 4 only comprised 2 and 3 cities, respectively. Here, Cluster 3 sticks out because of its high mean total population. Sticking with the population statistics, we see that Cluster 1 has a higher White population than the mean, while Clusters 2 and 3 are lower than the mean. The IoD and EI Z-score values imply that Cluster 1 cities are less segregated. This correlates with the medoid PIs from Figure \ref{fig:cubical_wbah_4_centroids}. We noted that Cluster 1 was the only one not containing Black, Asian, or Hispanic $H_0$ features born before $\varepsilon = 50$. This correlates with the fact that Cluster 1 cities are less segregated. The Black, Asian, and Hispanic Z-scores correspond to their respective IoD Z-scores.

\begin{table}
    \centering
    \resizebox{\textwidth}{!}{\begin{tabular}{c | R R R R R | R R R | R R R R}
      \bottomrule
       & \multicolumn{5}{|c} {\bf Population} & \multicolumn{3}{|c} {\bf IoD} & \multicolumn{4}{|c} {\bf EI}\\
      \multicolumn{1}{c} {\textbf{Cluster}} & \multicolumn{1}{|c} {Total} & \multicolumn{1}{c} {W} & \multicolumn{1}{c} {B} & \multicolumn{1}{c} {A} & \multicolumn{1}{c} {H} & \multicolumn{1}{|c} {W-B} & \multicolumn{1}{c} {W-H} & \multicolumn{1}{c} {W-A} & \multicolumn{1}{|c} {W-W} & \multicolumn{1}{c} {B-B} & \multicolumn{1}{c} {H-H} & \multicolumn{1}{c} {A-A} \\
      \toprule
      \bottomrule
      \textbf{1} & -0.16 & 0.26 & -0.15 & -0.05 & -0.05 & -0.17 & -0.25 & -0.17 & 0.11 & -0.14 & -0.11 & -0.10\\
      \textbf{2} & 0.08 & -0.78 & 0.46 & 0.19 & 0.07 & 0.36 & 0.60 & 0.43 & -0.36 & 0.38 & 0.24 & 0.31\\
      \textbf{3} & 5.32 & -0.81 & 0.17 & 0.28 & 0.51 & 1.24 & 1.35 & 0.98 & -0.27 & 0.50 & 0.69 & 0.76\\
      \textbf{4} & 0.29 & 0.01 & 0.10 & -0.42 & 0.40 & 0.83 & 1.06 & 0.35 & 0.11 & 0.36 & 0.63 & -0.29\\
      \toprule
    \end{tabular}}
    \caption{\label{table:wbah_cubical} Mean Cluster Statistics for $K$-Medoids White (W), Black (B), Asian (A), and Hispanic (H) Cubicical Clustering. The table's contents are the same as in Table \ref{table:wbah_levelset}, but we use WBAH cubical persistence instead.}
\end{table}

\subsection{Comparing Results from Different Clustering Methods}

Along with comparing certain statistics specific to each cluster, we also compare the clusters altogether. Introduced by William Rand in \cite{rand_index}, the Rand index measures the similarity between two data clusterings by iterating over all pairs of samples and counting the pairs assigned the same or different clusters between the two clusterings. After counting over all pairs, the Rand Index (RI) score is
\begin{align*}
    RI = \frac{\text{number of agreeing pairs}}{\text{number of pairs}}.
\end{align*}
Thus, an RI score of 0 means that the two clusterings do not agree at any point, while a score of 1 means that the clusterings are identical (up to a permutation of labels). While RI gives us a good understanding of the similarity between clusterings, we can further adjust for chance through the Adjusted Rand Index (ARI). The ARI score for two clusterings is given by
\begin{align*}
    ARI = \frac{\text{RI} - \text{expected RI}}{1 - \text{expected RI}},
\end{align*}
where the RI is the same as defined above, and the expected RI is the expected value of the RI if the clusterings are generated randomly. Thus, the ARI is always a score between $-0.5$ and 1, where a negative ARI score means a clustering is less similar to the other than a random cluster, and a score of $0$ means the clustering is just as good as random. A score of $1$ implies the clusterings are identical. Generally, a score of at least $0.5$ shows a good level of agreement between clusterings \cite{scikit-learn-rand}.

Using ARI scores to compare clusterings gives us an easy way of comparing many clusterings simultaneously. So far, we have only discussed clustering on White, Black, Asian, and Hispanic persistence data, but we do not have to use all four groups as input for clustering. For example, if we wanted to focus solely on White and Black segregation, we could omit the Asian and Hispanic persistence images from our clustering algorithm. While simultaneously considering level-set and cubical complexes, our number of clusterings grows significantly. Luckily, we can compare these clusterings through an ARI heatmap.

\begin{figure}[htbp]
    \centering
    \includegraphics[width=0.75\linewidth]{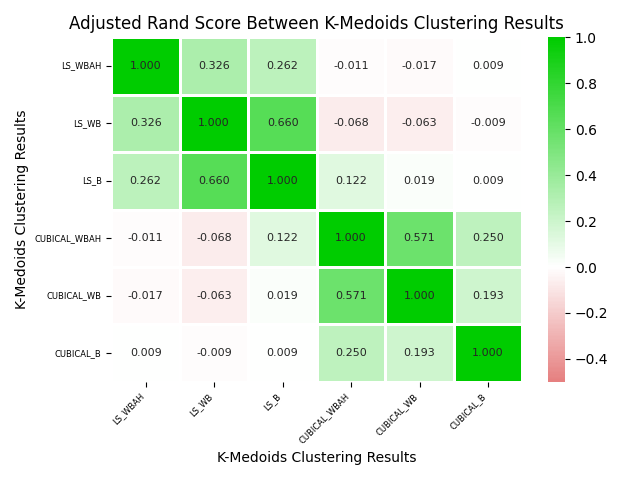}
    \caption{\label{fig:kmedoids_rand_index} A heatmap showing the Adjusted Rand Scores between the clustering results from the different persistence methods. Ranging from $-0.5$ to $1$, the ARI tells us how similar the two clustering results are.}
\end{figure}

Figure \ref{fig:kmedoids_rand_index} shows the ARI heatmap for six different $K$-medoids clusterings. Among these six clusterings, we consider three different sets of PIs using our two complex constructions: \{White, Black, Asian, Hispanic\}, \{White, Black\}, and \{Black\}. We notice a consistent similarity between the level-set and cubical clusterings. Specifically, \verb'LS_B' and \verb'LS_WBAH' have the least similarity of the level-set clusterings but are still moderately similar. The \verb'LS_B' and \verb'LS_WB' clusterings have an ARI score of 0.664, which shows a good level of agreement. Similar patterns can be seen for the cubical clusterings. Interestingly, the ARI between level-set and cubical clusterings seems to stay around $0$, which means that the agreement of the clusterings is no better than random.

\section{Conclusions and Future Work}
\label{conclusion}
Many methods for computing persistent homology require data to be in the form of a point cloud, which can lead to the loss of important geospatial information when areas are reduced to points. In this paper, we utilize level-set and cubical complexes as methods for persistent homology that preserve the geospatial features of our data. For a city-race pair $X-Y$, a level-set complex filtration is constructed by selecting the census tracts within city $X$ that have a majority population of race $Y$, then expanding the boundaries of these census tracts using the level-set method with a filtration parameter of ``time". A cubical complex filtration, on the other hand, is constructed by iterating over $\varepsilon$ where at each step, we select the census tracts in city $X$ that have a $Y$ population of at least $\varepsilon$ percent. Both of these approaches allow us to categorize the ``shapes'' of segregation in 112 cities across the United States by clustering the persistence images of our cities.

Within our methodology, we examine clustering results from our two complex construction methods and compare their features through medoid persistence images for each clustering. After obtaining our $K$-medoids clustering results, we also conduct clustering analysis by evaluating the mean cluster statistics for each method and determining the similarity of clustering results using the Adjusted Rand Index. Notably, the clusters exhibit distinct Index of Dissimilarity scores, indicating a correlation between our analysis and the Index of Dissimilarity. The Adjusted Rand Index analysis reveals high similarity within level-set clusterings and within cubical clusterings, but low similarity between the two methods.

There are many possible avenues for further work with our methods. While our analysis focused on $H_0$ and $H_1$ persistence for White, Black, Asian, and Hispanic demographic data, future studies could modify the features and data used. Our approach provides a general analysis of segregation, but it can be adapted to focus on specific types of segregation, such as Black-White segregation. Additionally, the analysis could be performed with different areal units. For example, using census blocks instead of tracts could facilitate a more detailed, neighborhood-to-neighborhood comparison within a city. Additionally, one could conduct temporal studies of persistent homology of geospatial data. While we only used 2020 Census data, future research could utilize past census data to examine how patterns of segregation have evolved over time.

Along with further applications of our methods, further explorations could compare level-set and cubical complexes of geospatial data in more depth. While we have compared the clustering results of the two methods, a more rigorous analysis of the methods themselves could provide valuable insights. As noted, level-set and cubical barcodes appear very different, with cubical barcodes containing significantly more features. Understanding these differences in greater detail could help explain the variations in clustering results. Additional future work could also investigate how choices in constructing level-set complexes influence the resulting persistent homology. For instance, as done in \cite{levelset}, we triangulate level-set images by selecting every fifth pixel as a vertex, but how does this choice affect the computed homology or computational complexity? Moreover, instead of triangulating, we could explore cubical persistence on these images, which may lead to different homological outcomes.

\section*{Acknowledgments}
We thank Laura Smith, Associate Professor of Geography at Macalester College, for helpful comments on our analyses. LZ was supported in part by grants from the
NSF (CDSE-MSS-1854703 and BCS-2318171).

\renewcommand{\bibfont}{\small}
\bibliographystyle{plainnat}
\bibliography{main.bbl}

\end{document}